\documentclass[journal,a4paper]{IEEEtran}

	\usepackage{citesort}
	\usepackage{subfigure}
	\usepackage{amsmath}\interdisplaylinepenalty=2500
	\usepackage{amsfonts}
	\usepackage{bbm}
	\usepackage{array}
	\usepackage{theorem}\newtheorem{remark}{Remark}
	\usepackage[pdftex]{graphicx}\graphicspath{{./FIGURES/}}

\usepackage{ifpdf}
\ifpdf
	\usepackage[pdftex,hypertexnames=false,colorlinks]{hyperref}
\DeclareGraphicsExtensions{.jpg,.pdf,.png,.tif,.fig} 
\else
\DeclareGraphicsExtensions{.eps,.eps.gz,.eps.Z,.jpg,.png,.fig} 
\fi

	\def\pth#1{\left(#1\right)}
	    
	\def\norm#1{\left\|#1\right\|}
	\def\cro#1{\left[#1\right]}

	\def\scal#1{\left<#1\right>}
	\def\scu#1{\ensuremath{\mathcal{#1}}}
	
	\def\nb{\boldsymbol{n}}
	\def\ib{\boldsymbol{i}}
	\def\jb{\boldsymbol{j}}
	\def\lb{\boldsymbol{l}}
	\def\ub{\boldsymbol{u}}
	\def\xb{\boldsymbol{x}}
	\def\vb{\boldsymbol{v}}

	\def\Gdeltap{\mathcal{G}_{\Delta'}}

	\def\percent{\%}

\input alphabeta.tex

\begin{document}

\title{An Improved Observation Model for Super-Resolution under Affine Motion}

\author{Gilles~Rochefort,~Fr\'ed\'eric~Champagnat,~Guy~Le~Besnerais,~and~Jean-Fran\c{c}ois~Giovannelli\thanks{G.~Rochefort is with Realeyes3D-France
217 Bureaux de la Colline - Rue Royale, Hall D - 92213 Saint-Cloud, France, E-mail:
grochefort@realeyes3d.com, F. Champagnat and G. Le Besnerais  are with
Office National d'Etudes et de Recherches A\'erospatiales (ONERA),
29 av. de la division Leclerc, BP 72, 92322 Chatillon Cedex,
France. E-mail:
firstname.lastname@onera.fr. J.-F.~Giovannelli
is with Laboratoire des Signaux et Syst\`emes, Sup\'elec, Plateau de Moulon,
3 rue Joliot-Curie, 91192 Gif-sur-Yvette Cedex, France. E-mail:
giovannelli@lss.supelec.fr. This work was performed at ONERA during the PhD of. G. Rochefort.}}

\markboth{An improved observation model for Super-Resolution under affine motion. Version of \today}{IEEE Transaction on Image processing (in revision)}


\maketitle

\begin{abstract}  
  Super-resolution (SR) techniques make use of subpixel shifts between
  frames in an image sequence to  yield higher-resolution images.
  We propose an original observation model devoted to the case
  of non isometric inter-frame motion as required, for instance, 
  in the context of airborne imaging sensors. First, we describe how the main 
  observation models used in the  SR literature deal with motion, 
  and we explain  why they are not suited for non isometric motion. 
  Then, we propose an extension of the observation model by Elad and
  Feuer 
 adapted to affine motion. This model is based on a
  decomposition of affine transforms into successive shear transforms, each
  one efficiently implemented by row-by-row or column-by-column 1-D affine
  transforms.
  
  We demonstrate on synthetic and real sequences that our observation
  model incorporated in a SR reconstruction technique leads to better 
  results in the case of variable scale motions and it provides 
equivalent results in the case of isometric motions.
\end{abstract}


\section{Introduction}

\PARstart{S}{uper-resolution} (SR) techniques aim at estimating a
high-resolution image with reduced aliasing, from a sequence of low-resolution
(LR) frames. The literature on the subject is abundant, see
\cite{super:tsai84,super:Schultz96,super:Patti97,Hardie97,Elad97,Farsiu04} and
\cite{Park03} for a recent review.

Our contribution deals with the class of ``Reconstruction Based'' SR
techniques~\cite{Baker02}, which can be split in three steps:
(1) estimation of inter-frame motion; 
(2) computation of a linear observation model including motion; 
(3) regularized inversion of the linear system.

We are interested in aerial imaging applications which often imply non
isometric motion, as in the case of an airborne imager getting closer to the
observed scene, see Sec.~\ref{sec:aerial}. Such non isometric motion fields
can be estimated using various registration
algorithms~\cite{Mann94,Thevenaz98}. Hence, step (1) is not the main issue in
this context. Concerning step~(2), 
the SR literature is rather allusive: most
published methods implicitly assume translational
motion~\cite{super:tsai84,Baker02,super:kim90,super:kim93,super:UretGross_improv_resol_sub_shift_pictur,Elad01,Hardie97,Farsiu04,Tom95,super:Tekalp92,Lee03,Patti01,Woods03}.
To the best of our knowledge, if some former contributions  
apply to affine
\cite{Hardie98,Elad99} or even homographic \cite{Mann94,Lertrattanapanich02}
warps none of them explicitly deals with variable distance from scene to imager
in step (2)\footnote{It is addressed formally in \cite{super:Patti97} but not
implemented nor demonstrated.}. So,we focus on the construction of a proper
observation model for affine motions with consistent scale changes. 

Section~\ref{sec:previous} proposes a bibliographical survey of the SR literature, with
respect to the observation model. It is shown that published methods are not
adapted to the considered context: its main difficulty is to account for
non translational motion in a tractable discrete model.

Section~\ref{sec:proposedmodel} is devoted to the proposed new 
observation model that extends
the popular one introduced by Elad and Feuer~\cite{Elad97} by replacing traditional
pointwise interpolation by techniques based on $\mathrm{L}_2$
approximations and shifted bspline basis~\cite{bspline:thevenaz9701}. 
We show that our model leads to a
more precise prediction of LR frame pixel values, in the case of a combined zoom
and rotation motion.

Further comparisons are performed on SR reconstruction results.
Section~\ref{sec:Inversion} briefly introduces the convex regularization
framework that we use for SR reconstruction. Such techniques are customary in
various inverse problems, including restauration and
SR~\cite{Idier01b,super:Schultz96,Elad97}.

We use the resulting SR reconstruction technique to compare various
observation models on synthetic (section~\ref{sec:synthetic}) and real
(section~\ref{sec:exper-results-real}) datasets. These experiments
consistently show that our model is more accurate and reliable for sequences
combining rotation and important scale changes, at the expense of 
a moderate increase of computational load.

\section{Analysis of previous works}
\label{sec:previous}

This section describes several published observation models 
differing by the way they account for motion through numerical approximations.

\subsection{Notations}\label{sec:notations}

Uppercase letters (resp. boldface letters) refer to matrices (resp. vectors).
$\nb =[n,l]^\mathrm{t} \in \mathbbm{Z}^2$ and $\ib = [i,j]^\mathrm{t} \in
\mathbbm{Z}^2$ denote discrete positions of LR and SR pixels and $\ub=[u,
v]^\mathrm{t} \in \mathbbm{R}^2$ denotes real positions on the image plane. An
image $x$ can be described by a continuous field $x(\ub)$, or by a sequence of
discrete coefficients $x\cro{\ib}$ and as lexicographycally ordered vector
$\xb$.

\subsection{General observation model}

Let $x\pth{.}$ be the input irradiance field and $y\cro{.}$ be the observed LR image. $y$ is a
sampled version of the convolution of $x$ with an optical point
spread function (PSF) $h_o$ integrated by a box function $I$ 
corresponding to the collecting surface of the detector:
\begin{equation*}
  y\cro{\nb}  =  \int_{\mathbb{R}^2}  \pth{h_{o}*x}\pth{\nb\Delta - \boldsymbol{v}} \, I\pth{\boldsymbol{v}} \, \mathrm{d}\vb \,, 
\end{equation*}
with $\nb \in \mathcal{G}_{\Delta}$. $\mathcal{G}_{\Delta} \subset \mathbbm{Z}^2$ is the set of
discrete detectors positions on a grid with step $\Delta$. Let us denote $N = \textrm{Card}\pth{\mathcal{G}_{\Delta} }$ the
number of LR pixels in frame $y$.

It is customary to define a joint optics-plus-detector PSF $h = h_o * I$ so
that $ y\cro{\boldsymbol{n}} = h*x\pth{\boldsymbol{n} \Delta}$.

SR methods rely on the usual ``brightness constancy''
assumption which is the basis of many motion estimation techniques, in
particular intensity-based techniques~\cite{Thevenaz98}. In this
framework, SR methods assume that temporally neighboring frames
originate from a unique input $x$ up to a warp  modeling relative
sensor\,/\,scene motion.

Let $y_k$ ($k=1,\dots,K$) denote a neighboring frame of $y$, then (\textit{i}) $y_k$
derives from an irradiance field $x_k$ through sensor $h$:
$y_k\cro{\boldsymbol{n}} = h*x_k\pth{\boldsymbol{n} \Delta}$ and
(\textit{ii}) there is a warp $w_k$, such that $x_k\pth{\boldsymbol{u}} = x\pth{w_k\pth{\boldsymbol{u}}}$.
%
%
Combination of both equalities yields
\begin{equation}\label{eq-directIdeal}
y_k\cro{\boldsymbol{n} } = h*\pth{x \circ w_k} \pth{\boldsymbol{n} \Delta} \,.
\end{equation}

The next step is discretization of $x$ for the sake of numerical computations. 
The irradiance field $x$ is decomposed on a shifted kernel basis:
\begin{equation}\label{eq-decompx}
x\pth{\ub } = \sum_{\ib\in\mathcal{G}_{\Delta'}} x[\ib] \, \varphi \pth{\ub - \ib \Delta'} \,.
\end{equation}
$\mathcal{G}_{\Delta'}$ is the SR grid, with step $\Delta'$ and $M =
\textrm{Card}\pth{\mathcal{G}_{\Delta'}}$ is the number of SR
pixels. The ratio $L = \Delta / \Delta'$ defines the practical
magnification factor (PMF) of the SR process: it is usually greater
than two. Note that it does not imply that the actual resolution
improvement is as high as the PMF. 

$\varphi$ may be any classical interpolation kernel (box function,
bilinear, ...). In the sequel, we use bspline basis, which
encompass most classical interpolation
schemes~\cite{bspline:CurrySchoenberg47,bspline:unser9301,bspline:unser9302}.
Then $\varphi$ is a separable bspline kernel of order $m$:
$\varphi\pth{\ub} = \beta^m\pth{u} \beta^m\pth{v}$, where $\beta^m(u)$
is the $(m+1)$-fold convolution of a box function.

Let us rewrite~(\ref{eq-directIdeal}) as:
\begin{equation}
   y_k\cro{\boldsymbol{n} } = \int_{\mathbb{R}^2}  x\pth{
   w_k(\vb)} h\pth{\nb\Delta - \boldsymbol{v}} \, \mathrm{d}\vb \, .
\label{eq-x-o-w-conv-h}
\end{equation}
Injecting~(\ref{eq-decompx}) yields:
\begin{equation*} 
y_k\cro{\nb} = \sum_{\ib\in\Gdeltap} a_k[\nb,\ib] \, x[\ib]  \,,
\end{equation*}
%
%
\begin{equation}\label{eq-defAni}
a_k[\boldsymbol{n},\boldsymbol{i}]  =  \int_{\mathbb{R}^2} \varphi \pth{w_k(\boldsymbol{v}) - \boldsymbol{i} \Delta'} \,
h(\boldsymbol{n} \Delta - \boldsymbol{v}) \, \mathrm{d}\boldsymbol{v} \,.
\end{equation}
Using lexicographically ordered vector representation of images, a matrix
formulation writes:
\[
  \yb_k = \AD_k \xb\,.
\]
The whole matrix $\AD = [ \AD_1 \ldots \AD_K ]^\mathrm{t}$ is huge with
dimensions $KN \times M$, $M\approx NL^2$. For instance, a sequence of $K=10$
frames, with dimensions $N=128^2$ and a PMF $L=2$ leads to about $43$ billion
elements. Of course, $\AD_k$ is a sparse matrix with a band structure, 
as practical PSF $h$ spreads over two or three LR pixels at most and 
$\varphi$ is a separable bspline kernel, whose support is 
$(m+1)\Delta'$ wide. However, the
cost of computing all non zero elements of $\AD$ remains formidable for
general warps $w_k$.

In the following, we review landmark SR papers with respect to 
the way they compute $\AD$. We discuss three main approaches:

\begin{enumerate}
\item Exact computation for special cases of $\wb_k, h$ and $\varphi$
\item \textit{Convolve-then-Warp} approximation
\item \textit{Warp-then-Convolve} approximation
\end{enumerate}


\subsection{Exact computation}\label{sec:exact-computation}

Exact computation is tractable only in two
special cases:
\begin{itemize}
\item motion is a global translation;
\item $\varphi$ and $h$ are both box functions and motion is affine.
\end{itemize}

\subsubsection{Global translation}
When $w_k$ is a
global translation, (\ref{eq-directIdeal})~leads to a simple
convolution. Indeed, replacing $w_k(\boldsymbol{u}) =
\boldsymbol{u} - \boldsymbol{\tau}_k$ inside~(\ref{eq-defAni}) yields:
\begin{equation*} 
a_k[\boldsymbol{n},\boldsymbol{i}]  = \varphi * h \pth{\boldsymbol{n} \Delta 
- \boldsymbol{i} \Delta' - \boldsymbol{\tau}_k} \,,
\end{equation*}
and the observation equation writes:
\begin{equation*} 
  y_k\cro{\boldsymbol{n}} = \sum_{\boldsymbol{i}}
  \varphi*h\pth{\boldsymbol{n}L\Delta' - \boldsymbol{i}\Delta' -
  \boldsymbol{\tau}_k} x\cro{\boldsymbol{i}} = g_k * x \cro{\nb L}
\end{equation*}
with $g_k(\ub) = (\varphi * h)(\ub\Delta' - \taub_k)$.
For a given integer $L$, each LR frame appears as a subsampled version of the
discrete convolution of $x$ with kernel $g_k$.

Most of the early SR literature is devoted to this global translation case. 
It naturally
leads to either Fourier
techniques~\cite{super:tsai84,super:kim90,super:kim93} or equivalent
multi-channel filtering
techniques~\cite{super:UretGross_improv_resol_sub_shift_pictur} based
on the generalized Papoulis theorem~\cite{Papoulis77}.

\subsubsection{$\varphi$ and $h$ are box funtions}
\begin{figure}[!htf]
  \centering
  \includegraphics[width=1.5in]{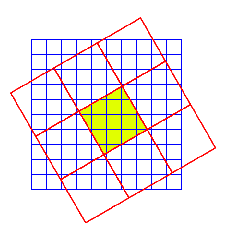}
  \caption{\small $\varphi$ and $h$ are assumed box functions and
  motion is a rotation.  
     The fine grid represents the grid of
    SR pixels, while the coarse one is the grid of detectors.  Common
    areas between the middle detector and each SR pixel are 
colored.\label{STARKfigs}}
  
\end{figure}

When $\varphi$ and $h$ are box
functions~\cite{super:Patti97,super:Schultz96,super:stark89},
(\ref{eq-defAni}) is the common area
between each detector and each warped SR pixel (see Fig.~\ref{STARKfigs}).

Such an observation model has been proposed by Stark and
Oskoui for rotational warps~\cite{super:stark89}.  No indication is
provided in their paper about the numerical computation of the
relevant intersections.

Assuming affine motion, each warped SR pixel is a convex polygon,
and computation of the intersection of two convex polygons
can be performed by a ``clipping'' algorithm
such as~\cite{sutherland74:Reentrant}. 
However, this technique is not suitable for SR purpose due to its high
computational burden.




\subsection{Convolve-then-Warp}\label{sec:convolve-then-warp}

Let us start back from~(\ref{eq-x-o-w-conv-h}). In practice, $h$
scarcely spreads over two or three LR pixels, thus integral
(\ref{eq-x-o-w-conv-h}) extends on a neighborhood
$\mathcal{V}\pth{\nb\Delta}$ around $\nb\Delta$.
Let us assume that
$w_k\pth{\ub}$ can be locally approximated by a translation:
\begin{equation*} 
w_k(\boldsymbol{u}) \approx w_k(\boldsymbol{n}\Delta) + \boldsymbol{u} - \boldsymbol{n}\Delta \,, \qquad \boldsymbol{u} \in \mathcal{V}\pth{\nb\Delta} \,.
\end{equation*}
Then (\ref{eq-x-o-w-conv-h})~can be approximated by a convolution:
\begin{equation} \label{eq:convolutionrelation}
  y_k\cro{\boldsymbol{n}} \approx   (h*x)(w_k(\boldsymbol{n}\Delta)) \,.
\end{equation}

\begin{figure}[!htf]
  \centerline 
  {
    \subfigure[Correct detector integration area.]
    {
      \includegraphics[width=1.8in]{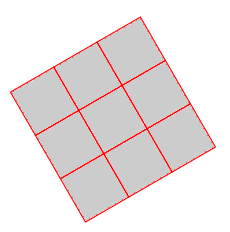}
      \label{SCHULTZTruefig}
    } \subfigure[Local translation approximation of the warp and
    resulting detector area.] {
    \includegraphics[width=1.8in]{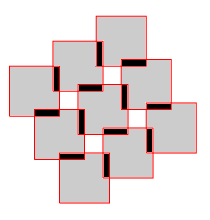}
      \label{SCHULTZRealfig}
    }
  }
  \caption{llustration of the Convolve-then-Warp approximate model~(\ref{eq:convolutionrelation}): 
  white regions are not accounted for,
  gray ones are integrated once while black regions are incorrectly
  integrated in two detectors output.\label{SCHULTZfigs}}
  
\end{figure}

Such an approximation is depicted in Fig.~\ref{SCHULTZfigs}. The
center of each detector is well positioned, but the integration
area is a rough approximation. Such an approximation leads to errors
in the integration step for large rotations and scale variations. 

The discretization of this model is much easier than the general
model~(\ref{eq-directIdeal}), because it is an irregular sampling of a
convolution. The simple model of Schultz and
Stevenson~\cite{super:Schultz96} is a special case of this approach
when $h$ and $\varphi$ are both box functions and the detector center
positions are rounded to integer multiples of $\Delta'$. Then, the
components $a_k[\boldsymbol{n},\boldsymbol{i}]$ are binary, with
$a_k[\boldsymbol{n},\boldsymbol{i}]=1$ if the $\boldsymbol{i}$-th SR
pixel is inside the $\boldsymbol{n}$-th detector area, approximated as in
Fig.~\ref{SCHULTZRealfig}. A refined version of this model is used
in~\cite{Irani91}.

As a conclusion, this model appears computationnaly attractive but is
clearly unable to correctly account for non-translational warps because
of the fixed detector geometry (see Fig.~\ref{SCHULTZfigs}).

\subsection{Warp-then-Convolve}\label{sec:warp-then-convolve}

This approach consists in using the convolution
relationship~(\ref{eq-directIdeal}) between the data
$\boldsymbol{y}_k$ and the warped SR image
$x_k(\boldsymbol{u})=x(w_k(\boldsymbol{u}))$. If a discretized version
$\hat{x}_k$ of $x_k$ over the $\Delta'$-shifted basis functions
$\varphi$ is available, (\ref{eq-directIdeal})~can easily be
discretized as:
\begin{equation*} 
  \boldsymbol{y}_k = \mathrm{DH}\boldsymbol{ \hat{x} }_k
\end{equation*}
where $\mathrm{D}$ is a down-sampling matrix, and $\mathrm{H}$ is the
convolution matrix associated to the optical-plus-detector response.

Now the main problem is to construct $\hat{\xb}_k$ using the
discretized SR image coefficients $x[.]$ defined by~(\ref{eq-decompx}). 
A first approach may be to enforce equality on the grid nodes:
\begin{equation*} 
  \sum_{\ib\in\Gdeltap} \hat{x}_k\cro{\ib} \varphi\pth{\pth{\lb-\ib}\Delta'}  =    \sum_{\jb\in\Gdeltap} x\cro{\jb} \varphi\pth{w_k\pth{\lb\Delta'} -\jb\Delta'} \,.  
\end{equation*}
If $\varphi$ is a bspline of order $m=0$ or $m=1$, it satisfies
$\varphi\pth{\pth{\lb-\ib}\Delta}=\delta\pth{\lb-\ib}$, and we get:
\begin{equation}\label{eq:expansionElad2}
  \hat{x}_k\cro{\lb} = \sum_{\jb\in\Gdeltap} x\cro{\jb} \varphi\pth{w_k\pth{\lb\Delta}  -\jb\Delta} \,.
\end{equation}

In other words, the discrete coefficient $\hat{x}_k\cro{\lb}$ is the
interpolation of $x$ at point $w_k\pth{\lb\Delta}$.  If $\varphi$ is a
box function ($m=0$), (\ref{eq:expansionElad2})~reduces to nearest
neighbor interpolation and if $\varphi$ is a triangle function
($m=1$), (\ref{eq:expansionElad2})~is a bilinear interpolation.

Interpolation~(\ref{eq:expansionElad2}) leads to the definition of a
warping matrix $\mathrm{W}_k$, which summarizes all motion
information. The complete image formation model is then:
\begin{equation}\label{eq:ELAD}
  \boldsymbol{y}_k = \mathrm{DHW}_k \xb \,.
\end{equation}
This is exactly the formulation proposed by Elad and
Feuer~\cite{Elad97,Elad99} referred to as ``E\&F'' model in the following.

\begin{figure}[!htf]
  \centering 
  \includegraphics[width=3in]{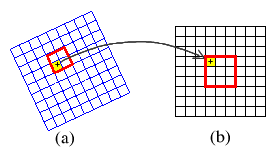}
  \caption{\small Illustration of the E\&F model: starting from SR image Fig.~\ref{ELADfigs}(a) an
    intermediate high-resolution image Fig.~\ref{ELADfigs}(b) is
    constructed with a pixel grid aligned with the $y_k$ data detector
    grid using either bilinear or nearest neighbor interpolation.\label{ELADfigs}}
  
\end{figure}

Fig.~\ref{ELADfigs} summarizes this method: starting from the sought
SR image Fig.~\ref{ELADfigs}(a), an intermediate high-resolution image
Fig.~\ref{ELADfigs}(b) is constructed with a pixel grid aligned with the
detector grid using either bilinear or nearest neighbor
interpolation. Integration and subsampling are then straightforward.

Compared to the previous approach, the
E\&F model seems much more precise for rotation warps. However, one
can foresee aliasing problems in the case of scale changes due to the
pointwise interpolation step~(\ref{eq:expansionElad2}). 

\section{Proposed observation model}\label{sec:proposedmodel}

This Section introduces an original observation model extending the
 E\&F model, by replacing pointwise
interpolation~(\ref{eq:expansionElad2}) by a technique based on
 $\mathrm{L}_2$ function approximation.


Dealing with variable scale using $\mathrm{L}_2$ approximation technique
is not easy in 2-D. In this context, Catmull and
Smith~\cite{catmull-80}  introduced an efficient decomposition of 2-D
affine transforms into separable 1-D transforms. 

First, we will introduce such decomposition into our observation
model. Next, we focus on the 1-D operations in order to
achieve a $\mathrm{L}_2$ approximation on a bspline basis. 
Finally, we will compare observation models and point
out the improvements provided by the proposed model.

\subsection{Warping decomposition}

\begin{figure}[!ht]
  \centering
  \begin{tabular}{cc}

  \subfigure[Original image.]{
    \includegraphics[width=3cm]{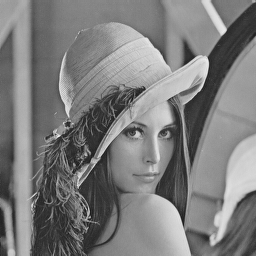}
  } &
  \subfigure[Horizontal shear.]{
    \includegraphics[height=3cm]{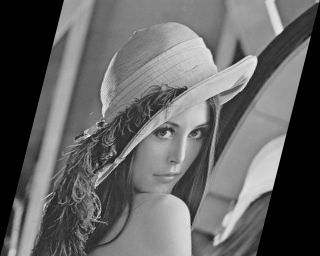}
  } \\
  &  \subfigure[Vertical shear.]{
    \includegraphics[height=2.436cm]{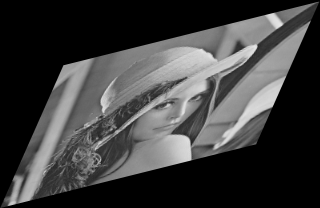}
  }
  \end{tabular}
  \caption{\small Example: the affine transform of
 ~(\ref{eq:exampleshear})  is decomposed in two steps. Each step 
    is  a shear along one coordinate image axe.\label{fig:cisaillement2pass}}
  
\end{figure}

Thevenaz and Unser showed that
2-D invertible affine transforms can be handled by 
two-shear or three-shear decompositions~\cite{bspline:thevenaz9701}. 
Each shear is a vertical or horizontal 
coordinate transform such as:
\begin{equation}\label{eq:elmtshearu}
  S_u(\ub) = 
  \begin{pmatrix}      \alpha_2 & \beta_2 \\ 0 & 1     \end{pmatrix}
  \begin{pmatrix}      u \\ v    \end{pmatrix} 
  +
  \begin{pmatrix}         \varepsilon_2 \\ 0       \end{pmatrix} \,,
\end{equation}

\begin{equation}\label{eq:elmtshearv}
  S_v(\ub) = 
  \begin{pmatrix}      1 & 0 \\  \beta_1 & \alpha_1  \end{pmatrix}
  \begin{pmatrix}      u \\ v    \end{pmatrix} 
  +
  \begin{pmatrix}       0 \\  \varepsilon_1       \end{pmatrix} \,.
\end{equation}
Both ot them are one-dimensional affine transforms separably applied 
row-by-row or column-by-column. 
As an example, Fig.~\ref{fig:cisaillement2pass} provides the
intermediate images resulting of each shear of the following affine motion and decomposition:
\begin{equation}\label{eq:exampleshear}
  \begin{pmatrix}
    1 & 1/4 \\  -1/4&  7/16
  \end{pmatrix} =   \begin{pmatrix}
    1 & 0 \\  -1/4&  1/2
  \end{pmatrix}
  \begin{pmatrix}
    1 &  1/4\\ 0 & 1 
  \end{pmatrix} \,.
\end{equation}

This decomposition is not unique, and the choice of one particular
decomposition impacts the transformed image quality. Catmull and
Smith~\cite{catmull-80} mentioned the bottleneck problem resulting
from a down-scaling in one pass followed by up-scaling in the next
pass, resulting in a loss of resolution.

Many approaches have been proposed to minimize image degradation,
depending on the considered transform. For intance,
Paeth~\cite{multipass:paeth86} has proposed a three-shear
decomposition well-suited for rotation. Other authors refer to
$N$-pass decomposition~\cite{multipass:fraser94}.

Multi-pass interpolation techniques and their limitations are outside
the scope of this article, the reader can refer
to~\cite{multipass:fraser94} for deeper insight. In the sequel, we
consider only two-shear decompositions. In this case, there are two
possibilities, and one selects the decomposition which reduces the
involved scale
variations~\cite{bspline:thevenaz9701,bspline:horbelt0101,bsplines:munoz0101}.
%


\subsection{1-D affine transform approximation}

Let us consider a 1-D affine transform with parameters $(a, \tau)$: $f(u)
\rightarrow f\pth{(u - \tau)/a}$. With this notation, $a<1$ yields a signal
reduction and $a>1$ yields a signal magnification. It is clear that signal
reduction may result in important discretization errors (as naive subsampling
undergoes a frequency aliasing).

In the line of Thevenaz \textit{et al.} \cite{bspline:thevenaz9701}, let us decompose $f$ on the 1-D shifted bspline basis:
\begin{equation} \label{eq:decf}  
  f\pth{u} = \sum_{k\in\mathcal{G}_Q} f\cro{k} \beta^m\pth{u - k} \,,
\end{equation}
where $\mathcal{G}_Q \subset \mathbbm{Z}$ denotes the set of $Q$
discrete samples (for instance the set of pixels of a row of the image).
We search for coefficients $g\cro{k}$, $k\in\mathcal{G}_Q$ such
that $g$, defined by
\begin{equation} \label{eq:decg}   
  g\pth{u} = \sum_{k\in\mathcal{G}_Q} g\cro{k} \beta^m\pth{u - k}  \,,
\end{equation}
achieves the best approximation of $f\pth{(u - \tau)/a}$
in the $\mathrm{L}_2$ sense, \textit{i.e.} minimization of $\int \cro{
  f\pth{ (u - \tau)/a}-g(u) }^2 \mathrm{d}u$. The 
approximation is the orthogonal projection, and the optimal coefficients
satisfy the orthogonality equations
\begin{equation} \label{eq:projectederror}
  \scal{  g\pth{u} -
    f\pth{\frac{u-\tau}{a}},\,\beta^m\pth{u-k}}  =  0\,,
\end{equation}
for $k\in\mathcal{G}_Q$. Replacing~(\ref{eq:decf}) and~(\ref{eq:decg}) in~(\ref{eq:projectederror}) yields:
%
%
\begin{equation*} 
  \sum_j g\cro{j} \beta^{2m+1}\cro{j-k}  =  \sum_l f\cro{l} a\,\xi^{m}_a(k-\tau-al) \,,
\end{equation*}
with $\beta^m_a\pth{u} = \beta^m\pth{u/a} / a$ and $\xi^{m}_a=\beta^m_a * \beta^m$. The so-called
bi-kernel $\xi^{m}_a$ encodes the geometric
transform of a sample to a different scale space~\cite{bsplines:munoz0101}, and actually
provides an optimal anti-aliasing filter~\cite{bspline:unser9504}.
If $a\neq 1$, $\xi^{m}_a$ is not a bspline kernel, but remains a piecewise polynomial.
A closed form expression of $\xi^{m}_a$ 
is provided in~\cite{bspline:horbelt0101}.

Finally, the sought coefficients $g\cro{k}$ write:
\begin{equation} \label{eq:relationcoeffs}
   g\cro{k} = \pth{\beta^{2m+1}}^{-1} *
\left(a\,\sum_{l\in\mathcal{G}_Q}
   f\cro{l} \xi^{m}_a(k-\tau-al) \right) \,,
\end{equation}
and the inverse filter $\pth{\beta^{2m+1}}^{-1}$ can be
efficiently implemented through recursive filtering~\cite{bspline:unser9302}.

To sum up the process, given a sequence of signal samples $f\pth{k}$
and 1-D affine transform parameters $(a, \tau)$ the approximation goes
through four steps: 
\begin{enumerate}

\item compute bspline coefficients $f\cro{k}$; 
\item compute the bi-kernel function $\xi^{m}_a$; 
\item compute $g\cro{k}$ with~(\ref{eq:relationcoeffs}) and 
\item post-filter coefficients $g\cro{k}$ to get samples values $g\pth{k}$.
\end{enumerate}

\begin{remark}---
The first and the last steps are not required when the bspline
representation order $m$ is $0$ or $1$. Indeed, for these particular
orders, bspline coefficients are identical to image samples.
\end{remark}
\begin{remark}--- \label{rq-translation}
In the translation case ($a=1$), $\xi^{m}_a (u)= \beta^{2m+1} (u)$.
The $L_2$ approximation then turns to a mere bspline interpolation 
with a higher-order kernel.  
\end{remark}

\subsection{A two-shear observation model}

In the proposed model, the $k$-th observed frame $\boldsymbol{y}_k$ (in vector notation) writes:
\begin{equation*} 
  \boldsymbol{y}_k = \mathrm{DHS}_k^1\mathrm{S}_k^2\mathrm{{\boldsymbol{x}}} \,,
\end{equation*}
where $\mathrm{S}_k^1$ and $\mathrm{S}_k^2$ are shear operators. Each operator is
an 1-D row-by-row (or column-by-column) affine transform, which is implemented
as described in the previous section. In the sequel, we use an order-0 bpline
kernel. Thus, as a consequence of Remark 2, our model is identical to that of
Elad and Feuer with bilinear interpolation for translation motion.
The resulting model is denoted TS0 for Two-Shear model with 0-order bspline
basis.

\subsection{Comparing observation models}\label{compallmodels}

In this section we illustrate the quality of each observation model compared
to exact computation in the special case of $h$ and $\varphi$ chosen as box
functions and affine motion, see Sec.~\ref{sec:exact-computation}.

\begin{figure}[htbp]
  \subfigure[scale factor $1$.]
  {
    \includegraphics[width=0.9\columnwidth]{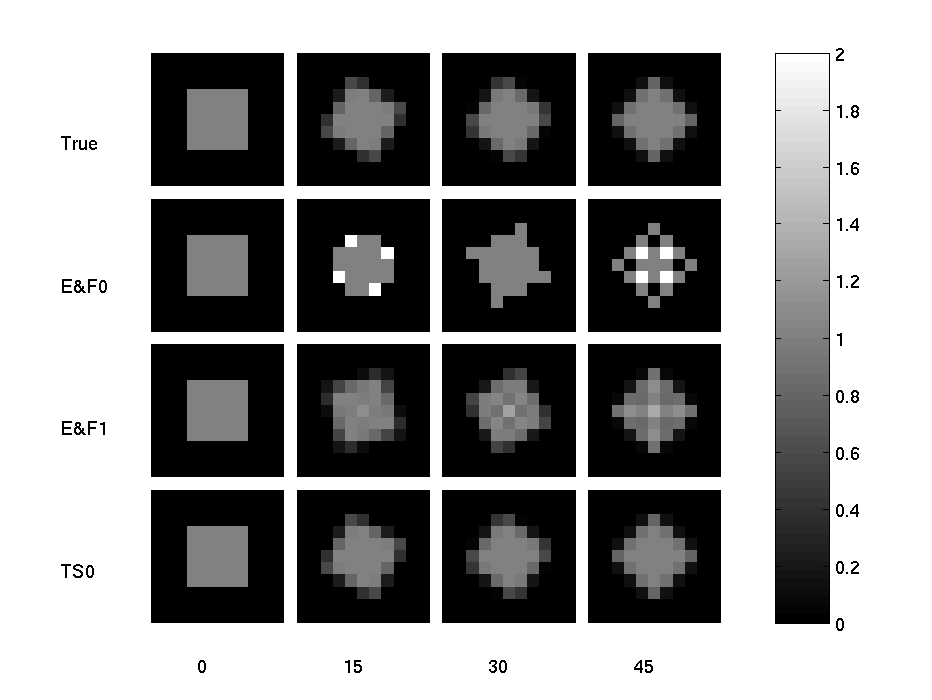}
    \label{coeff1}
  }
  \subfigure[scale factor $1.2$.]
  {
    \includegraphics[width=0.9\columnwidth]{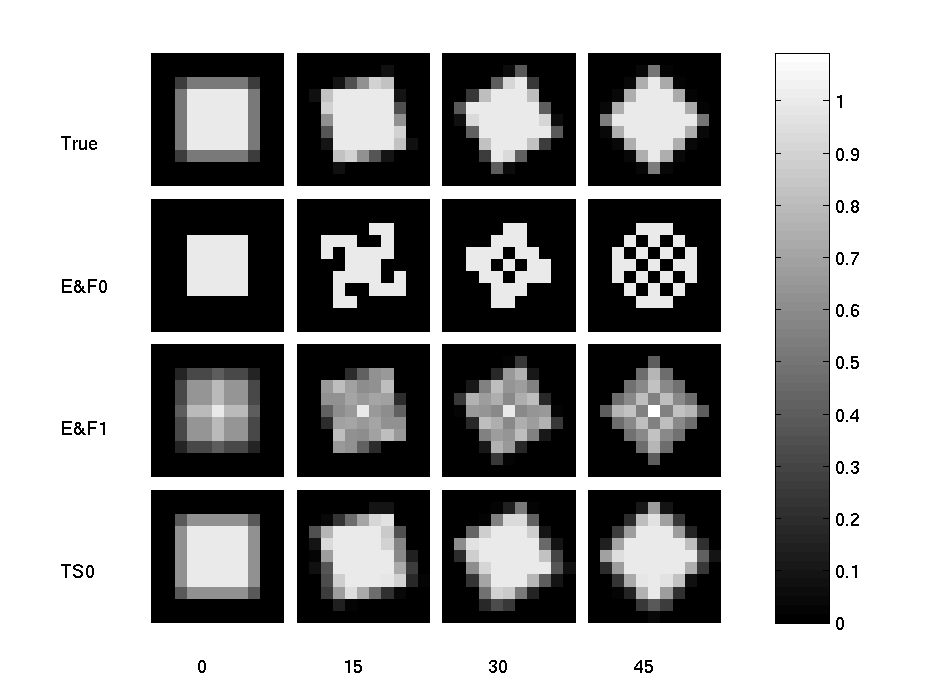}
    \label{coeff12}
  }
  \subfigure[scale factor $1.6$.]
  {
    \includegraphics[width=0.9\columnwidth]{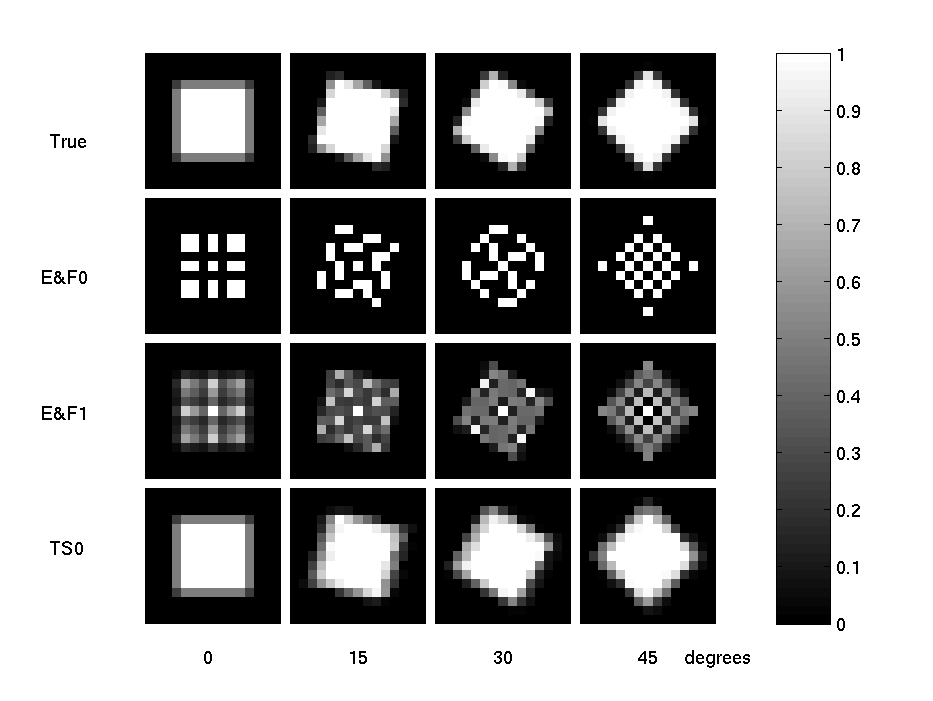}
    \label{coeff16}
  }
  \caption{\small Comparing observation models: SR pixels contributions to
    one detector. Scale factor $1.0$ \ref{coeff1}, $1.2$ \ref{coeff12} and 
    $1.6$ \ref{coeff16}, rotation up to 45 degrees. The models being 
    compared come from the 
    E\&F methods with order $0$ (E\&F0) and order $1$ (E\&F1) interpolation.
    The last line shows the proposed TS0 model, while the first
    line shows the true contributions. \label{Contribs}}
  
\end{figure}

We represent the components of the observation matrix $a_k\cro{\nb,\bullet}$
for a unique LR pixel in the form of an image patch. This patch displays the
weighting coefficients actually applied on SR image pixels for computing one
LR detector output. The first rows of the following three arrays of patches
show the exact components for rotation angles $\{0, 15, 30, 45\}$ degrees,
scale variations of~$1$ (Fig.~\ref{coeff1}), $1.2$ (Fig.~\ref{coeff12}) and
$1.6$ (Fig.~\ref{coeff16}) and a PMF of $5$.
 
The remaining patches show the approximated components
obtained using Elad and Feuer models with nearest neighbor
interpolation (E\&F0) or with bilinear interpolation (E\&F1) and the
proposed model (TS0).

The Convolve-then-Warp model is not presented, but would lead to the same
image patch made of a fixed size square pattern, whatever rotation and zoom
factor.


Fig.~\ref{Contribs} shows that E\&F0 is always incorrect even with limited
rotations and/or scale variations. It is noticeable that in Fig.~\ref{coeff1},
some coefficients value reach two: some SR pixels (white colored) contribute
twice to the detector. Such a behavior has been previously observed for the
``Convolve-then-Warp'' approach, see Fig.~\ref{SCHULTZRealfig}. In the same
time several SR pixels do not contribute at all to the detector.

\begin{figure}[htbp]
    \includegraphics[width=\columnwidth]{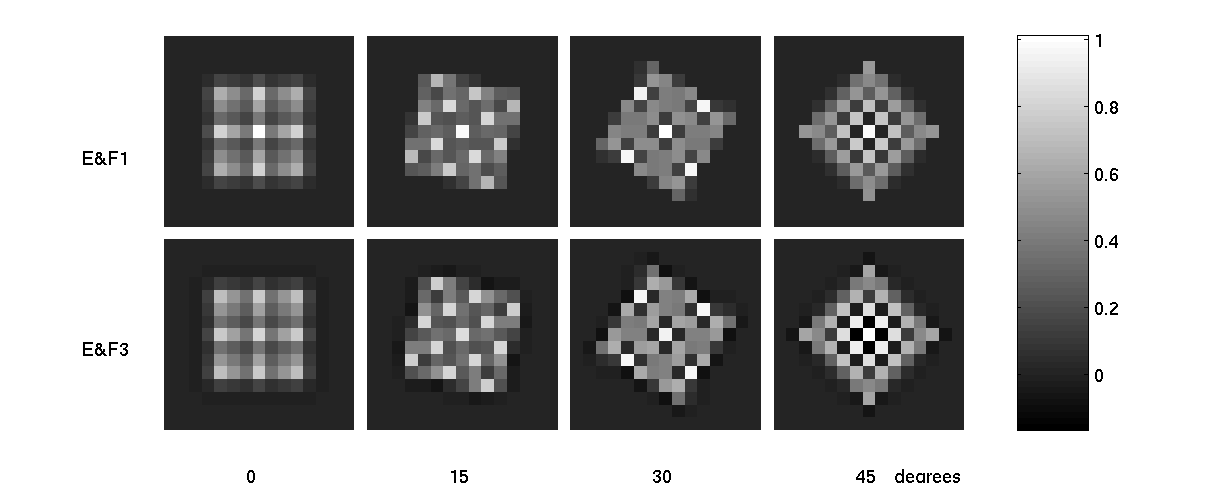}
  \caption{\small Comparing E\&F1 model with an Elad and Feuer model with
  bicubic interpolation (E\&F3), Scale factor 1.6 and rotation up to 45
  degrees.\label{ContribsEF3}}
  
\end{figure}

E\&F1 provides a better approximation. Still, contributions of SR pixels are
not uniform inside the detector footprint. This is already observed in
Fig.~\ref{coeff1} with rotations, and takes more importance in
Fig.~\ref{coeff12} and Fig.~\ref{coeff16} with scale factor and
rotations. 
As E\&F1 contributions appear as a smoothed version of E\&F0 ones, 
one wonders if a bicubic interpolation
(E\&F3) would give correct contributions. This is not the case, as shown by
Fig.~\ref{ContribsEF3}. Moreover, as bicubic interpolation does not preserve
positivity, the E\&F3 model exhibits negative contributions.

Whatever the interpolation method, Elad and Feuer models become inaccurate
for rotations as low as $15^\textrm{\,o}$ and zooming factor as low as
$20\percent$.

In contrast, the TS0 observation model ensures that the 
contributions of SR pixels are uniform inside the detector footprint
 whatever rotation and/or
scale factor being applied. Remaining differences between exact contributions
and TS0 ones are located on the detector boundaries: 
TS0 contributions spread on
slightly more than true ones.

\section{Regularization framework}\label{sec:Inversion}

The inversion step is tackled within a classical convex regularization
framework~\cite{Idier01b} as in many other SR 
methods~\cite{super:Schultz96,Elad97}. The estimated SR image is the
(possibly constrained) minimizer of a regularized criterion based on
observation model and convex edge-preserving penalty:
\begin{equation}\label{eq:convexgeneral}
J_\lambda\pth{\boldsymbol{x}} = \sum_k 
	\norm{\boldsymbol{y}_k - \AD^{\sdm {model}}_k\boldsymbol{x}}^2  + \lambda \sum_{c \in \mathcal{C}} \psi_s\pth{\boldsymbol{v}_c^t \boldsymbol{x}} \,.
\end{equation}
The first term of criterion~(\ref{eq:convexgeneral}) is a least squares
discrepancy between data and model output: $\AD^{\sdm {model}}_k$ stands for
the observation model which is to be inverted and derives either from the 
Elad and
Feuer approach or from the proposed model of Sec.~\ref{sec:proposedmodel}. The
second term is a convex penalization term~\cite{Idier01b}. $\mathcal{C}$ is the
set of cliques: it consists of all subsets of three adjacent pixels either
horizontal, vertical and diagonal. $\boldsymbol{v}_c$ denotes a second-order
difference operator within clique $c$. The regularization parameter $\lambda$
balances the trade-off between the two terms of the criterion. The potential
$\psi_s$ is chosen as a $\mathrm{L}_2-\mathrm{L}_1$ hyperbolic function:
\begin{equation*} 
  \psi_s\pth{u} = 2 s \pth{ \sqrt{ s^2 + u^2 } - s } \,. 
\end{equation*}
Parameter $s$ sets the threshold between the quadratic behavior ($u\ll s$),
which allows small pixel differences smoothing and the linear behavior
($u\gg s$) aimed at preserving edges. The latter part produces a lower
penalization of large differences compared to a pure quadratic function.
$\psi$ has the same qualitative behaviour as the Huber function of \cite{super:Schultz96}.

Finally, for a given observation model, four solutions are computed, based on:
\begin{itemize}
\item quadratic penalty
\item quadratic penalty and positivity constraint
\item hyperbolic penalty
\item hyperbolic penalty  and positivity constraint.
\end{itemize}

The criterion is convex by construction and has a unique global minimizer. The optimization can be achieved by iterative gradient-like techniques~\cite{nocedal99} and we resort to a limited memory BFGS algorithm\footnote{The implementation named VMLMB, has been provided by \'Eric Thi\'ebaut (thiebaut@obs.univ-lyon1.fr).}. It belongs to the class of Quasi-Newton algorithms which only requires evaluation of the criterion and its gradient (no second order derivative is explicitly needed) and it is known to have better convergence properties than gradient algorithms. 


\section{Experiments with synthetic sequences} 
\label{sec:synthetic}

This section presents the experiments conducted on
synthetic sequences. Using synthetic sequences has two main advantages:
\begin{itemize}

\item Sequences are built from a reference HR image which will later
  be used as a reference to compare with reconstructed SR images;
  
\item We control all imaging parameters: noise level, PSF and image size. Motion is exactly known too.

\end{itemize}

\subsection{Synthetic data}\label{synthese-seq}

To generate a sequence of LR frames, the observation matrices $\AD_k$ are
computed exactly according to assumptions of
Sec.~\ref{sec:exact-computation} that $\varphi$ and $h$ are box functions. As
previously said, such a technique is very time consuming.

\begin{figure}[!htf]
  \centering
  \subfigure[Lena.]
  {
    \centering
    \includegraphics[width=1.1in]{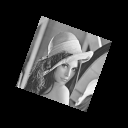}
    \includegraphics[width=1.1in]{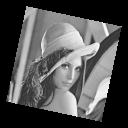}
    \includegraphics[width=1.1in]{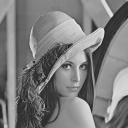}
    \label{fig:lenafirstlast}
  }

  \subfigure[Mire.]
  {
    \centering
    \includegraphics[width=1.1in]{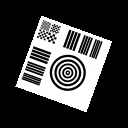}
    \includegraphics[width=1.1in]{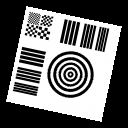}
    \includegraphics[width=1.1in]{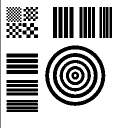}
    \label{fig:mirefirstlast}
  }
  \caption{\small We show the first, middle and last 
    frame of sequences \emph{Lena}~\ref{fig:lenafirstlast} and
    \emph{Mire}~\ref{fig:mirefirstlast}.}
\end{figure}

We simulate a smooth motion with a rotation up to $20$ degrees and
a zoom factor up to $1.6$. Each frame is $128 \times 128$ and is built from a $256
\times 256$ HR reference image. In Fig.~\ref{fig:lenafirstlast}, we show the
first, middle and last frame generated from the reference HR image
\emph{Lena}.

We also generate another sequence from a bitonal calibration pattern
named \emph{Mire}. The first, middle and last image of the sequence are
shown in Fig.~\ref{fig:mirefirstlast}.

\subsection{Results}\label{sec:Results}

Four regularized solutions and three observation models (E\&F0, E\&F1 and TS0)
are then available.
%
Hence, we finally compare performances of 12 SR settings with respect
to the reference HR image, by means of the PSNR (Peak Signal-to-Noise
Ratio, PSNR$=20 \log_{10}\pth{ {255}/{\sqrt{e}}}$, with $e$ the
mean square error). For each setting, the presented result is obtained
with the best regularization parameter (\textit{i.e.}, selected to
get the highest reachable PSNR).

\begin{figure}[!htf]
  \centering
  \subfigure[\small No additional noise.]
  {
    \includegraphics[width=3in]{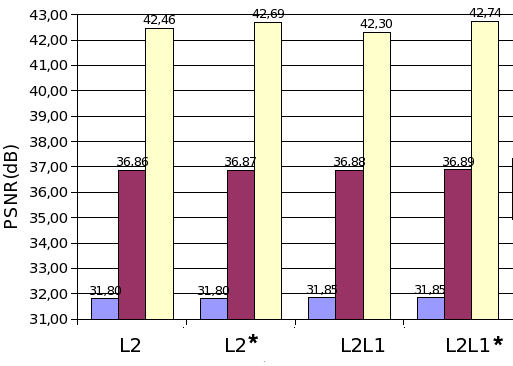} 
    \label{fig:psnrgraphlenasb}
  }  
  \subfigure[\small Additive Gaussian noise of variance $2$.]  {
    \includegraphics[width=3in]{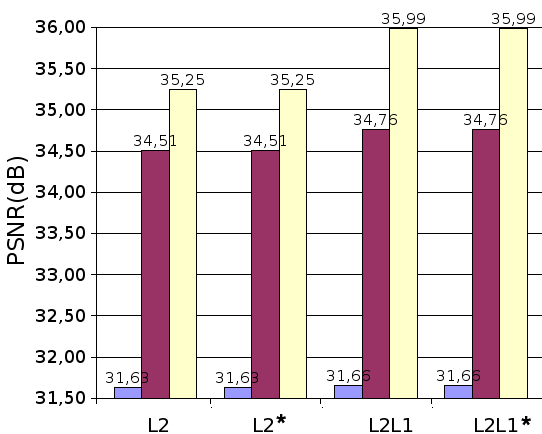}
    \label{fig:psnrgraphlenab2}
  }
  \caption{ \small SR performances on the \emph{Lena} sequence. Three
    observation models (E\&F0 (cyan), E\&F1 (magenta) and TS0 (yellow)) and four criteria are compared. Solutions
    which use a positivity constraint are labelled with a star. }
  
\end{figure}

Let us first deal with the ``Lena'' sequence of
Fig.~\ref{fig:lenafirstlast}. Fig.~\ref{fig:psnrgraphlenasb} sums up
the performance levels which have been achieved. First note that, on
these relatively smooth images, various regularization settings lead
to similar performances, and unconstrained quadratic regularization
suffices to obtain good results.
But we observe strong differences between observation models.
On the average, there is an improvement from $4$ dB (noisy case) up to $6$
dB (no noise) between the E\&F0 and the E\&F1 models. Moreover, there is also a gain
of $1$ to $6$ dB between the E\&F1 model and the TS0 model.

\begin{figure}[!ht]
  \centering
  \subfigure[\small No additional noise.]
  {
    \includegraphics[width=2.75in]{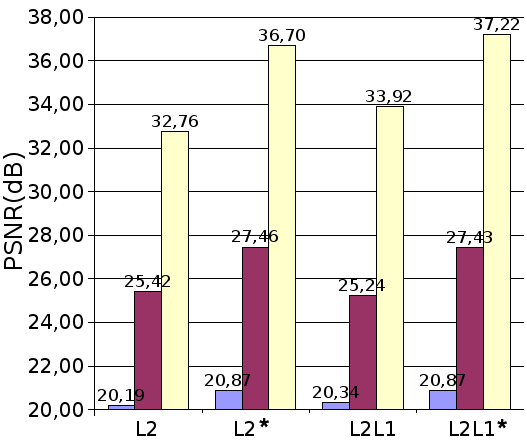} 
    \label{fig:psnrgraphmiresb}
  }
  \subfigure[\small Gaussian noise of variance $2$.]
  {
    \includegraphics[width=2.75in]{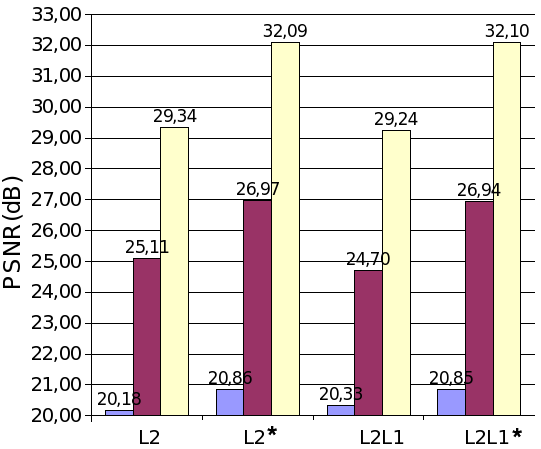} 
    \label{fig:R3}
  }

  \caption{ \small SR Performances on sequence \emph{Mire}. Three observation
    models (E\&F0 (cyan), E\&F1 (magenta) and TS0 (yellow)) and four criteria
    are compared. Positivity constraint is labelled with a star. \label{fig:psnrgraphmires}}

\end{figure}

\begin{figure*}[!htb]
  \centerline
  {
    \subfigure{ \includegraphics[width=1.8in]{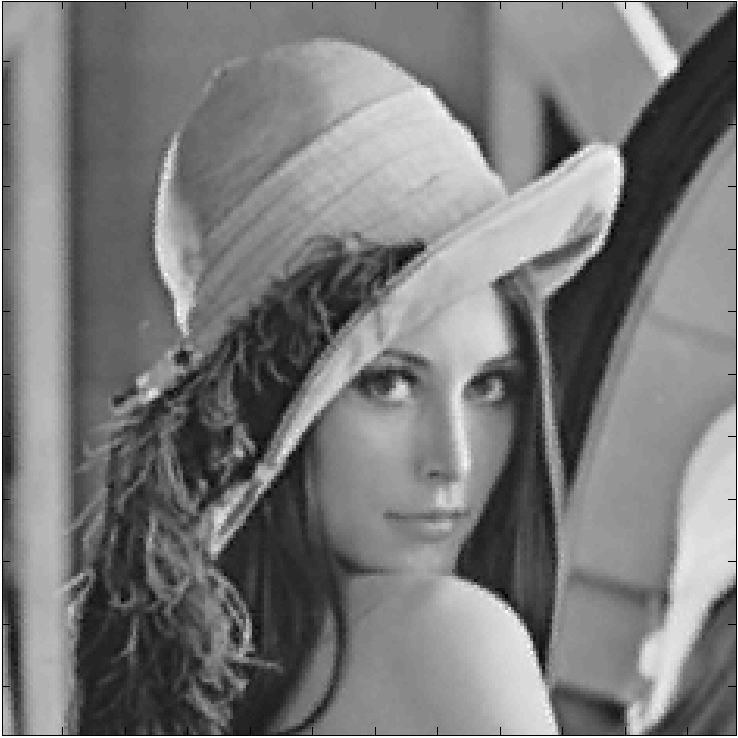} }
    \subfigure{ \includegraphics[width=1.8in]{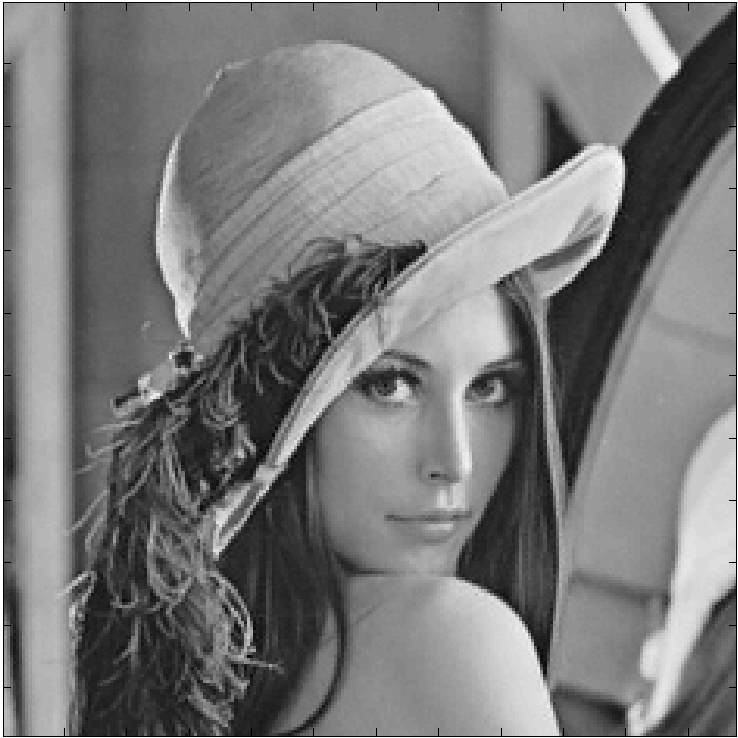} }
    \subfigure{ \includegraphics[width=1.8in]{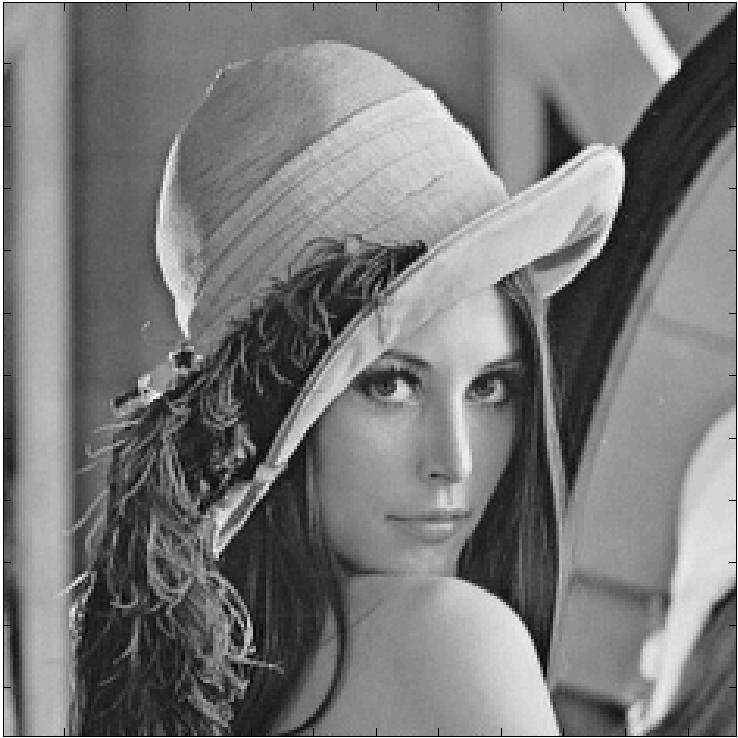} }
  }
  \centerline
  {
    \subfigure{ \includegraphics[width=1.8in]{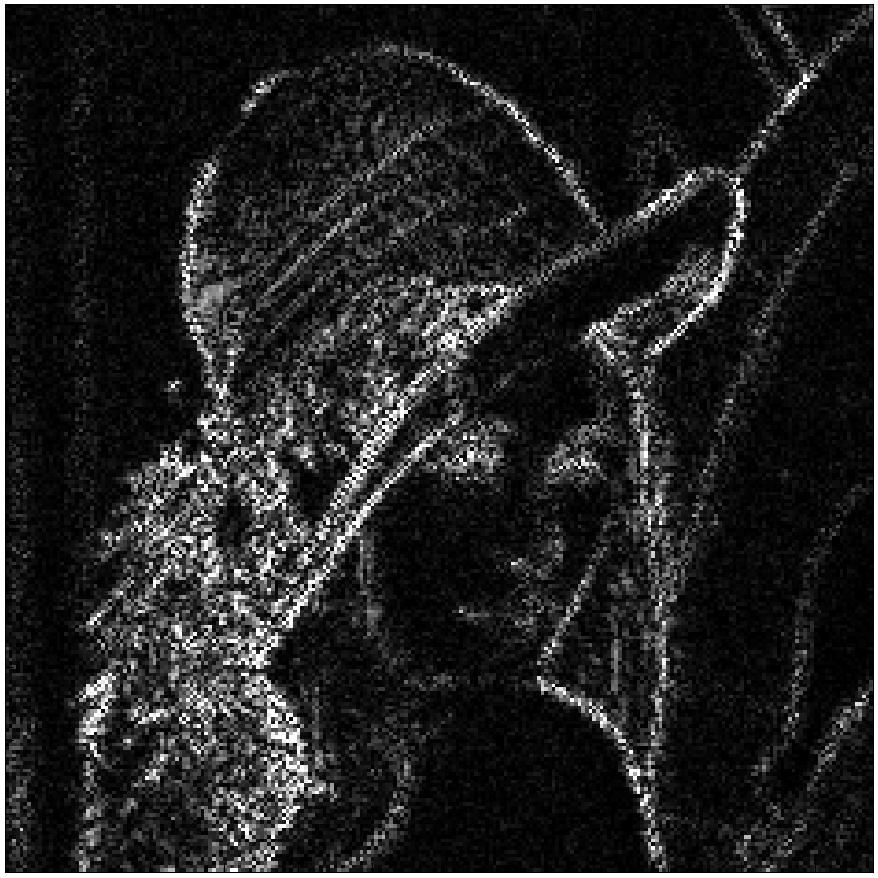} }
    \subfigure{ \includegraphics[width=1.8in]{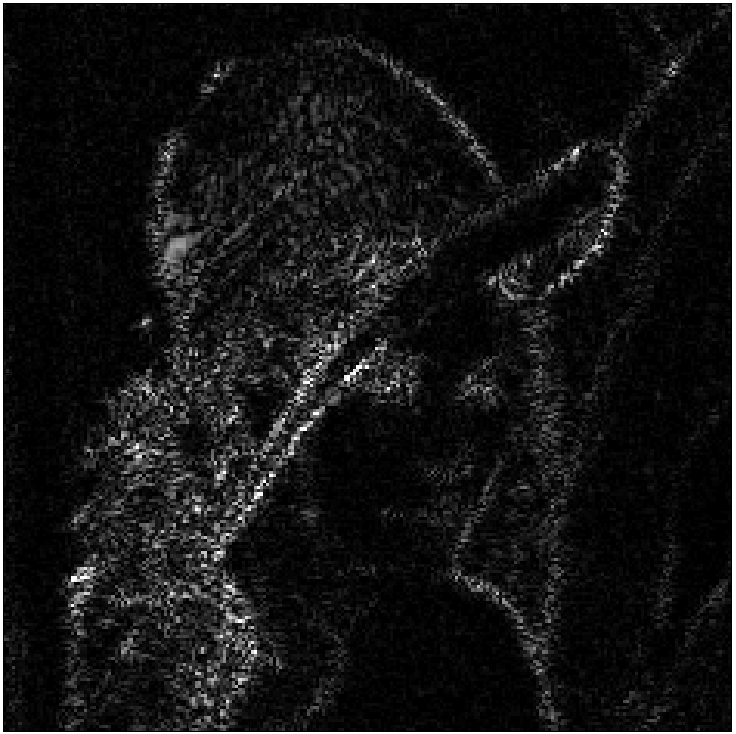} }
    \subfigure{ \includegraphics[width=1.8in]{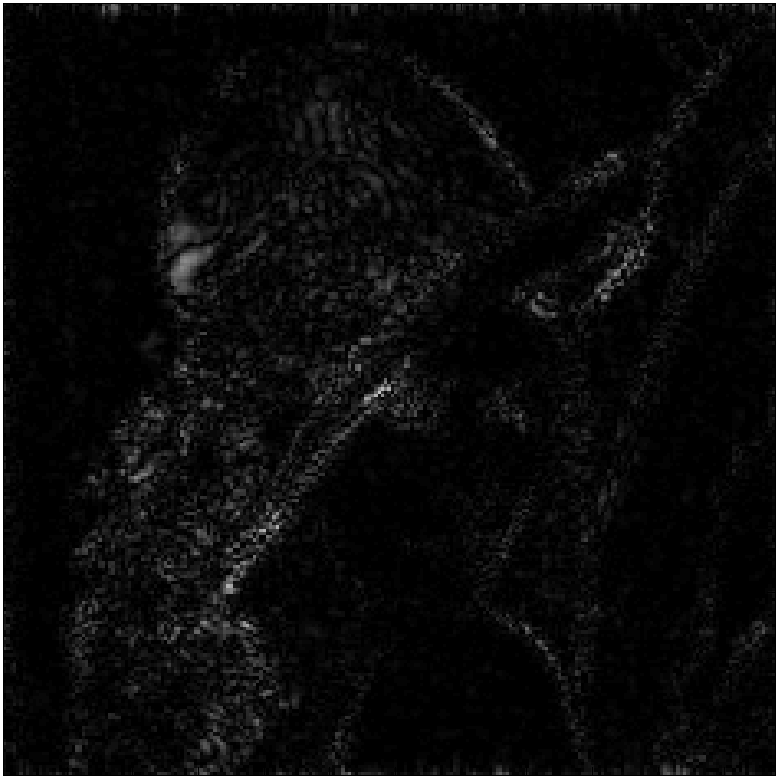}     }
  }   
  \caption{ \small First row, reconstructed SR images. From left to right:
    E\&F0, E\&F1 and TS0 observation model. All reconstructions are performed
    with a hyperbolic regularization and positivity constraint. Second row:
    differences between HR reference image and reconstructed SR images.\label{fig:seq1}  }
  
\end{figure*}

Fig.~\ref{fig:seq1} illustrates the differences between reconstructed
SR images, using $\mathrm{L}_2-\mathrm{L}_1$ regularization and positivity constraint,
depending on the chosen observation model. Once again, the
reconstructed images shown on the first row of Fig.~\ref{fig:seq1}
have been obtained with the best regularization parameters. The E\&F
reconstructions are slightly more blurred than the SR image obtained
from the proposed TS0 model. This is confirmed in the lower row which
shows image error with respect to the reference HR image: the TS0
observation model yields a better reconstruction on high frequency
areas, like the feather on the hat or the eyes.

We have also measured CPU time on a Pentium 4 at 2.66GHz. For this particular
sequence, one iteration duration is respectively $2.0$ and $4.6$ seconds, for
E\&F0 and E\&F1 methods. Our model requires $5.9$ seconds per iteration. All
methods converge roughly with the same number of iterations. Hence our method
is $30\%$ more time consuming than E\&F1.

We now consider the bitonal ``Mire'' sequence shown in
Fig.~\ref{fig:mirefirstlast}.  Results are reported in
Fig.~\ref{fig:psnrgraphmires} in terms of PSNR. As expected, this
high-frequency sequence leads to much stronger differences between
regularization terms and constraints.

\begin{figure*}[!htb]
  \centerline{
    \subfigure[E\&F0.]
    { 
      \includegraphics[width=1.8in]{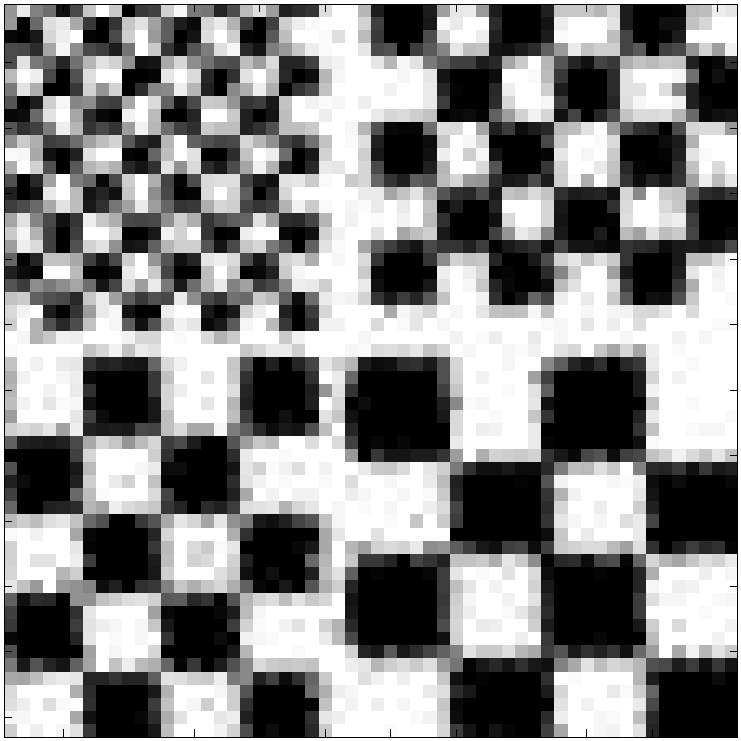} 
      \label{fig:R2_E0}
    }
    \subfigure[E\&F1.]
    {
      \includegraphics[width=1.8in]{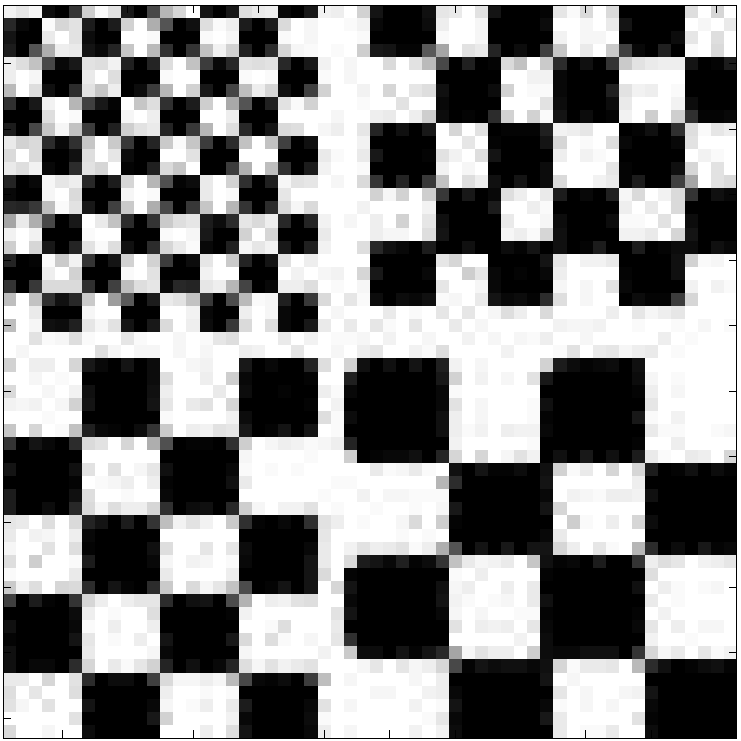}  
      \label{fig:R2_E1}
    }
    \subfigure[TS0.]
    {
      \includegraphics[width=1.8in]{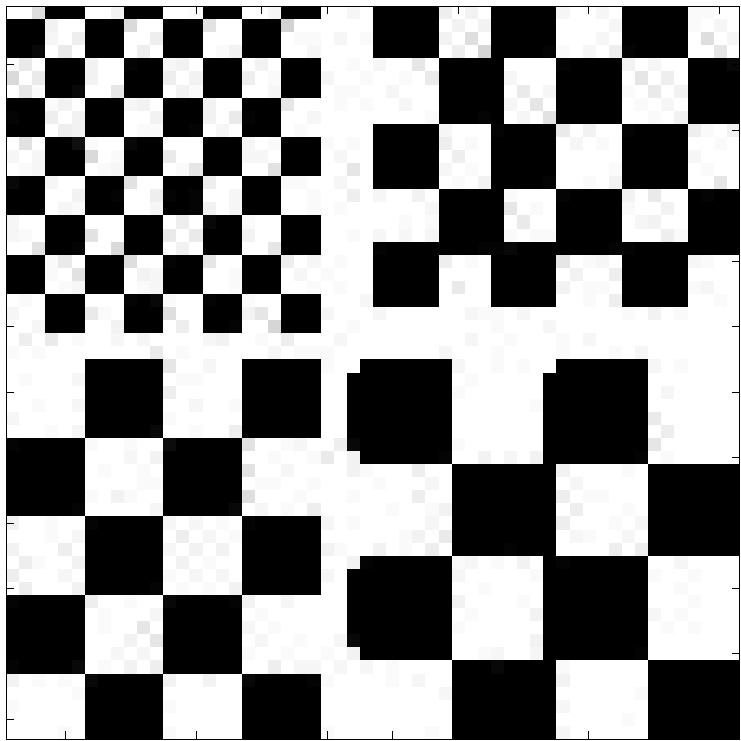}  
      \label{fig:R2_S0}
    }    
  }
  \caption{ \small Top-left parts of SR reconstructed images with hyperbolic
    regularization and positivity constraint. \ref{fig:R2_E0}: E\&F0 model,
    \ref{fig:R2_E1}: E\&F1, and \ref{fig:R2_S0}: proposed TS0 model.
    Parameters have been adjusted to get the best PSNR w.r.t. HR reference
    image.\label{fig:R2}}
  
\end{figure*}

As previously, strong differences are observed between observation models. On
the average, there is a gain improvement from $5$ dB (noisy case) up to $10$
dB (no noise) between E\&F1 model and TS0. Such an improvement is due to the high
contrast in \emph{Mire} image. Indeed, we know from Sec.~\ref{compallmodels}
that our observation model does not induce non homogeneous contributions in
the case of variable scale motion. The induced errors in the 
reconstructions are more 
visible in high contrast areas, see Fig.~\ref{fig:R2} compared to Fig.~\ref{fig:seq1}.

We also note that, in the noiseless case, hyperbolic regularization does not
improve performances of E\&F methods, whereas we notice a gain up to $1$ dB on
the average, with the TS0 model.

E\&F reconstructions are much more noisy than the one obtained with the
TS0 model. Let us recall that these reconstruction are obtained with a
regularization parameter adjusted to get the better PSNR w.r.t. the reference
HR image. The selected regularization parameter is lower ($10^{-4}$) with the
TS0 model than with E\&F models ($10^{-3}$). This might indicate that the more
precise the model is the less it is necessary to regularize. In other words,
regularization compensates for model errors which are lower with the proposed
TS0 model.

By using synthetic sequences with rotational and variable scale motion, we
have shown that the TS0 observation model leads to better reconstructed SR
images than E\&F methods, whatever the regularization involved.

As a general comment, it should be emphasized that performances are much more
sensitive to a change of observation model than to a change of
regularization.
In other words,  a good choice of the observation model leads to much higher
improvement than changing the regularization term, 
at least in the context of rotation and scale variation explored here.
\section{Experiments on real sequences}\label{sec:exper-results-real}

In this section, we compare observation models on real sequences. We first
discuss prior assumptions on the sequences with an emphasis on motion
modelization and estimation, then we present the results obtained on two real
datasets.

\subsection{General assumptions and motion estimation}

SR requires the knowledge of the sensor response and of the motion field between
frames. We use the common box function model for the PSF. Note that all the tested
observation models can accomodate a more general PSF.

We restrict our experiments to affine motion between frames, since the
proposed TS0 model is limited to these motion fields. Affine model accurately
describes the motion of a planar scene through orthographic
projection~\cite{Mundy92}. Such assumptions are usually not valid on the whole
field of view (except in special purpose experiments, see~\ref{sec:labTest}),
nevertheless the affine motion model is often a good local approximation of
complex motion fields~\cite{Mann94}, valid in a restricted part of the image
support (see an example in the aerial sequence of Sec.~\ref{sec:aerial}).

We focus on sequences which exhibit large affine motions, with total zoom
factor greater than $1.4$ and rotations higher than $20$ degrees (with inter-frame
zoom up to $1.2$ and rotation $5$ degrees). Note that such experimental settings
are not considered in the previous papers on SR, even those which address the
non translational context~\cite{Mann94,Lertrattanapanich02}.

The first problem is to register each image of the sequence with respect
to the reference image (usually the more resolved one). In this context,
direct intensity based methods, which minimize a DFD (displaced frame
difference) criterion are subject to false local minima, even using a
multiresolution approach. This is due to the sensitivity of DFD criterion with
respect to large rotational and scale changes. Hence, we use a two-step
approach:
\begin{enumerate}
\item compute a rough affine motion from scale-invariant keypoints matching;
\item refine the affine model using multiresolution DFD minimization on
  a restricted part of the image.
\end{enumerate}
The first step uses Scale-Invariant Fast Transform (SIFT) keypoints of D.
Lowe~\cite{Lowe04}. We match hundreds of keypoints between the considered
frame and the reference one by SIFT descriptor correlation, then we robustly
fit an affine model on selected matches using a crude rejection threshold. The
second step is essentially a domestic version of the pyramidal image
registration method of Thevenaz \textit{et al.}~\cite{Thevenaz98}.


\subsection{Lab tests}
\label{sec:labTest}

\begin{figure}[htbp]
\begin{center}
  \includegraphics[width=0.4\columnwidth]{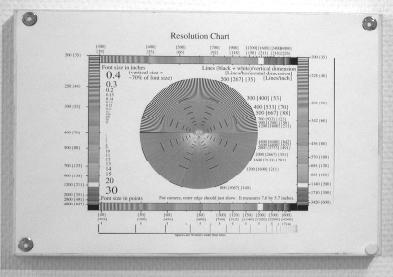}
  \includegraphics[width=0.4\columnwidth]{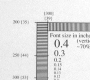}\\
  \includegraphics[width=0.4\columnwidth]{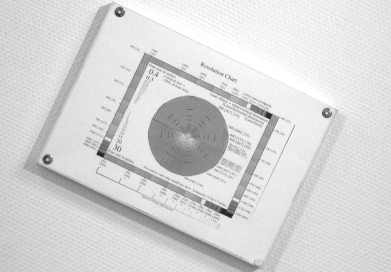}
  \includegraphics[width=0.4\columnwidth]{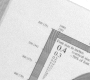}
  \caption{\small A sample of frames of the resolution chart, for various
  rotations and zoom factors, left column shows a zoom on the region used for
  further SR comparison. Up: reference frame, which is the most 
resolved one. \label{fig:expeLab1}}
 
\end{center}
\end{figure}

We made several SR experiments by using sequences of a bitonal resolution
chart printed on an A4 paper sheet observed with a AVT-046B SVGA Marlin B/W
camera. We acquired image sequences with variable inter-frame translation,
rotation and zoom factor: some examples are shown in
Fig.~\ref{fig:expeLab1}. Each frame of a sequence is registered with respect
to the reference frame as explained in the previous section. We ran SR
reconstructions with the three concurrent observation models and quadratic or
hyperbolic regularization, subject to positivity constraint. For each setting,
several values of the regularization parameter have been tried. Indeed, most
of the time there is a certain range of (low) values of the parameter where
differences between methods can easily be observed, whereas above some
regularization strength, all methods become equivalent and yield an
oversmoothed result.

\begin{figure}[htbp]
\begin{center}
  \includegraphics[width=0.9\columnwidth]{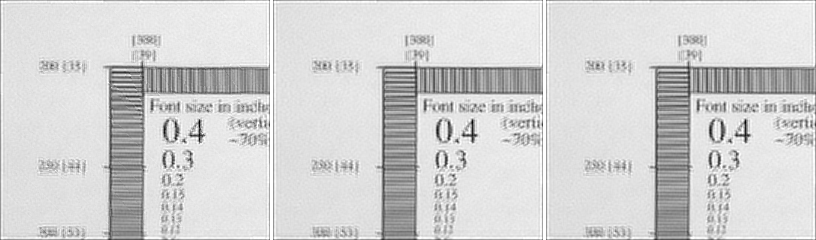}\\
  \caption{\small Reconstruction results with PMF $L=3$ using $7$ frames with global
  translation motion, in an under-regularized quadratic setting,
  $\lambda=7.10^{-3}$. From left to right: E\&F0, E\&F1 and TS0 models.\label{fig:mireLabT}}
  
\end{center}
\end{figure}

As a first example, we process a purely translational sequence, using $7$ frames
with a PMF $L=3$ and a quadratic regularization: comparison on a small
($240\times240$) region is shown in figure~\ref{fig:mireLabT}, for a low value
of $\lambda=7.10^{-3}$. As expected, in this case E\&F1 and TS0 lead to
quasi-identical results (PSNR = 68dB) whatever the parameter $\lambda$, while
E\&F0 shows some instability for low $\lambda$.

\begin{figure}[htbp]
\begin{center}
  \includegraphics[width=0.9\columnwidth]{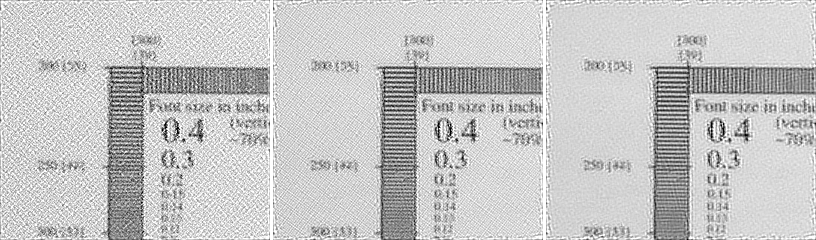}\\
  \includegraphics[width=0.9\columnwidth]{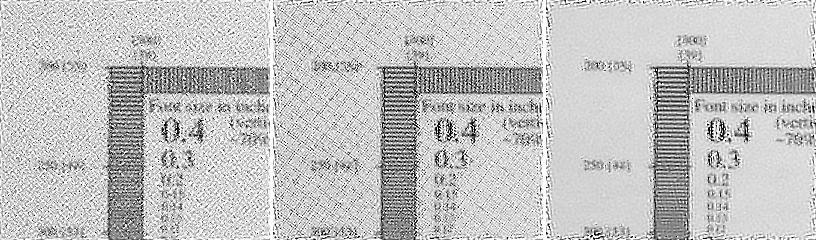}
  \caption{\small Reconstruction results with PMF $L=3$ using $7$ 
frames with zoom
  and rotations, in an under-regularized setting. Up: quadratic
  regularization, $\lambda=10^{-3}$. Down: hyperbolic regularization, $s=10$,
  $\lambda=3.10^{-3}$. From left to right: E\&F0, E\&F1 and TS0 models.\label{fig:mireLabZR1}}
  
\end{center}
\end{figure}

\begin{figure}[htbp]
\begin{center}
  \includegraphics[width=0.9\columnwidth]{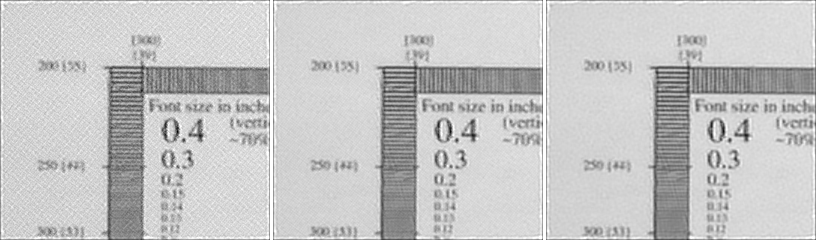}\\
  \includegraphics[width=0.9\columnwidth]{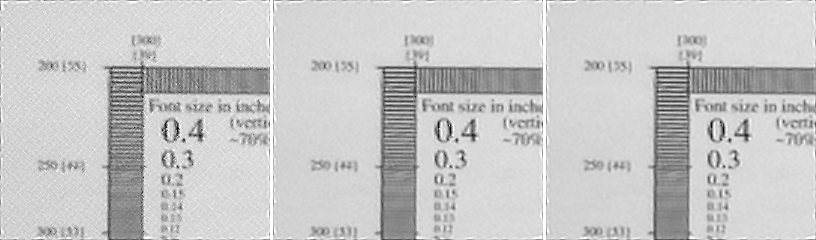}
  \caption{\small Reconstruction results with PMF $L=3$ using $7$
 frames with zoom
  and rotations, using a balanced regularization strength. Up: quadratic
  regularization, $\lambda=10^{-2}$. Down: hyperbolic regularization, $s=10$,
  $\lambda=3.10^{-2}$. From left to right: E\&F0, E\&F1 and TS0 models.\label{fig:mireLabZR2}}
  
\end{center}
\end{figure}

Fig.~\ref{fig:mireLabZR1} and Fig.~\ref{fig:mireLabZR2} show compared SR results on
7 frames of a sequence with both rotation (up to $25$ degrees) and zoom (there is
a factor $1.5$ between the reference image and the farthest view). We use either
quadratic regularization (upper part of the figures) or hyperbolic
regularization with a threshold parameter $s=10$ (lower part).

For a low value of the regularization parameter ($\lambda=10^{-3}$ with
quadratic term and $\lambda=3.10^{-3}$ with hyperbolic regularization), see
Fig.~\ref{fig:mireLabZR1}, E\&F0 and E\&F1 suffer from artifacts in the form of
a pseudo-periodic texture, which is of high amplitude in E\&F0 and less
important, but manifest, in E\&F1. Not surprisingly, this phenomenon is
amplified by the hyperbolic regularization. For the same regularization
parameter, TS0 does not encounter such instabilities, but exhibits ripples
which are typical of an under-regularized quadratic solution, 
and appear amplified by the hyperbolic edge-preserving potential.

For a more balanced value of the regularization parameter, see
Fig.~\ref{fig:mireLabZR2}, E\&F0 is still clearly degraded by instabilities.
E\&F1 and TS0 are now very close, but a careful examination of both
solutions reveals that small amplitude artifacts remain in the E\&F1
reconstruction.

\subsection{Aerial sequence}
\label{sec:aerial}

\begin{figure}[!htb]
  \centering
  \subfigure[First frame.]
  { 
    \includegraphics[width=0.9\columnwidth]{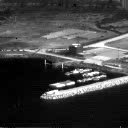}
  }
  \subfigure[Last frame.]
  { 
    \includegraphics[width=0.9\columnwidth]{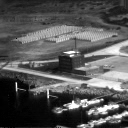}
  }  
  \caption{\small IR sequence captured by an airborne sensor, motion resuts
from variable distance and small rotation.\label{fig:seqreel}}
  
\end{figure}

Fig.~\ref{fig:seqreel} displays the first and the last frames of an infrared
sequence captured by an array sensor mounted on an airborne platform. As the
plane gets closer to the scene, the last frame is the most resolved one and is
chosen as the reference frame.
The scene is a harbour with the sea and waterfront in the foreground, a
building with a vertical antenna in the middle and a series of cans lined up
in the background. Two ships are present in the right low part of the last
frame. Because of perspective effects -- the lowest part of the frame is
closer to the sensor than the upper part -- apparent motion is closer to an
homography than an affinity. From the first frame to the reference one, the lower part
(resp. upper part) of the field of view is magnified with a factor about $1.4$
(resp. $1.6$). Therefore our method can only be applied to small regions of the
frames.

\begin{figure}[!htb]
  \centering 
    \includegraphics[width=0.9\columnwidth]{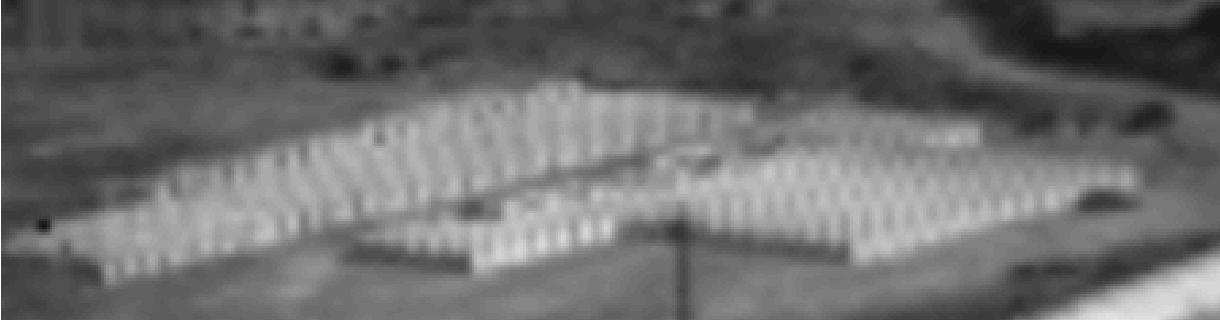}
  
  \caption{Detail of the last (reference) frame. Lined-up cans zoomed up twice
    using bilinear interpolation. The cans are not resolved. The black
    vertical line in the low middle of the image is the antenna on the
    building seen in Fig.~\ref{fig:seqreel}.\label{fig:details} }
\end{figure}

Two regions are considered in the sequel: \textit{(i)} in the upper part of the scene,
the lined-up cans that remain unresolved in the reference frame (see Fig.
\ref{fig:details}) and \textit{(ii)} in the right low part of the scene, the waterfront
and the ships, see Fig.~\ref{fig:SRresultsBateauxInterp}.

We considered five frames of the sequence, Fig.~\ref{fig:seqreel} displays two
of them. As already described, motion is estimated using SIFT on the whole
sequence then the intensity based method of \cite{Thevenaz98} is used to
refine the SIFT estimate in each region.

SR reconstruction is performed with the algorithms of Sec.~\ref{synthese-seq},
with quadratic regularization ($s = \infty$) and positivity constraint.
PMF $L= 2$ along both image axis.

\subsection{Upper region}

\begin{figure}[!htb]
  \centering 
    \includegraphics[width=0.9\columnwidth]{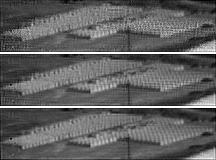}
 
  \caption{Reconstructions obtained through
    E\&F0 (top image) , E\&F1 (middle image) and TS0 (bottom image) 
observation model. 
$\lambda = 5. 10^{-3}$. \label{fig:SRresultsBidonOK}}
  
\end{figure}

The observation models are compared through the SR reconstructions in
Fig.~\ref{fig:SRresultsBidonOK}.

\begin{figure}[!htb]
  \centering 
    \includegraphics[width=0.9\columnwidth]{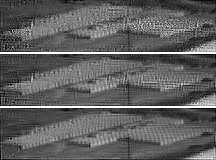}
 
  \caption{Reconstructions obtained through E\&F0 (top image) , E\&F1 (middle
    image) and TS0 (bottom image) observation model. $\lambda = 1. 10^{-3}$. \label{fig:SRresultsBidonSous}}
  
\end{figure}

The image quality in Fig.~\ref{fig:SRresultsBidonOK} gradually increases from
the top image (E\&F0) to the bottom image (TS0 model). Even if the latter is
still not a high quality image, the improvement in resolution enables the
count of the right block of cans in the bottom image, whereas it is less
obvious in the middle image and even impossible in the upper image. The
results of Fig.~\ref{fig:SRresultsBidonOK} look somewhat oversmooth. So a
lower regularization parameter has been tested, results are displayed in
Fig.~\ref{fig:SRresultsBidonSous}.

Fig.~\ref{fig:SRresultsBidonSous} reveals that E\&F0 and E\&F1 
are severely affected by the decrease of the regularization parameter,
whereas our model seems more robust: artifacts appear in the 
right top part of the scene, but cans can still be counted.

\subsection{Right lower region}

\begin{figure}[!htb]
  \centering 
    \includegraphics[width=0.9\columnwidth]{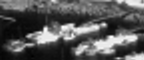}
 
  \caption{Detail of the last frame of Fig.~\ref{fig:seqreel}. Low
    right part of the scene: waterfront and ships zoomed up
    twice using bicubic interpolation.\label{fig:SRresultsBateauxInterp}}
  
\end{figure}

\begin{figure}[!htb]
  \centering 
    \includegraphics[width=0.9\columnwidth]{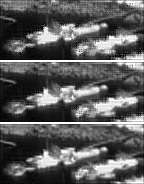}
 
  \caption{ \small Reconstructions have been performed using
    E\&F0 (top image), E\&F1 (middle image)  and TS0 
(bottom image) observation model.
 $\lambda = 10^{-2}$. \label{fig:SRresultsBateaux}}
  
\end{figure}

Fig.~\ref{fig:SRresultsBateaux} proposes similar results for 
the ships at the right low part of the scene. The ships appear
in bright contrast. A bicubic interpolation 
of the last observed frame is provided
in Fig.~\ref{fig:SRresultsBateauxInterp}. 
The top image (E\&F0 model) in Fig.~\ref{fig:SRresultsBateaux} has many
localized high frequency artifacts, part of them are absent in the middle
image (E\&F1 model). These artifacts are not present in the bottom image
(proposed TS0 model). In the same time, comparison of SR results and
Fig.~\ref{fig:SRresultsBateauxInterp} shows that resolution has indeed been
increased.

\section{Conclusion}


The presented paper deals with SR techniques in the field of aerial imagery.
The proposed work focuses on the observation model in the case of an affine
motion whereas the main part of SR literature deals with the inversion
process or motion estimation.

We analyzed the existing observation models used in SR reconstruction and
emphasized their underlying assumptions, so as to clarify their limitations.
As a result, it is shown that these observation models fall into three
categories:
\begin{itemize}
\item exact computation
\item convolve-then-warp
\item warp-then-convolve
\end{itemize}
Exact computation is not tractable for general motions. The
convolve-then-warp approach is numerically efficient but is unable to capture
large rotations and scale variations. So, only the third approach, due to Elad
and Feuer is relevant in our framework. However, we have observed inaccuracies
for rotations as low as $15^\textrm{\,o}$ and zooming factor as low as
$20\percent$.
We succeeded in extending the E\&F model to cover a more important range of
affine transforms with high accuracy, for about $30\percent$ more
computation time. The pointwise interpolation stage in the E\&F method has been
replaced by $\mathrm{L}_2$ functional approximation techniques. This technique
combines a two-shear decomposition for the affine transform and a 1-D
$\mathrm{L}_2$ projection on a shifted bspline basis.

The proposed model has been compared with various E\&F-like models. These
models have been associated to several regularization settings to be tested
for SR reconstruction purposes using synthetic and real image sequences.

These tests have stressed the importance of the observation model in SR
reconstruction when dealing with large zoom and rotation effects. In
particular the choice of a bilinear interpolation instead of a
nearest-neighbor one within an Elad and Feuer setting dramatically improves
the reconstructions. Moreover, the proposed model consistently achieves even
better results.

Further research should be conducted to accurately deal with homographic
motion, or piecewise parametric motion. It should unlock SR techniques to a
larger application field.

\section*{Acknowledgment}
The authors would like to thank \'Eric Thi\'ebaut for providing an
implementation of the VMLMB algorithm (see Sec.~\ref{sec:Inversion}), used
to optimize the constrained regularized  criteria.


\begin{biography}[{\includegraphics[width=1in,height=1.25in,clip,keepaspectratio]{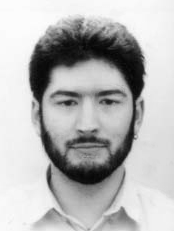}}]{Gilles Rochefort}
was born in Issy les Moulineaux, 
France, in 1977. He graduated from the \'Ecole Sup\'erieure de M\'ecanique 
et d'\'Electricit\'e in 2000. He received the Doctorat
degree in signal processing at Universit\'e Paris-Sud, 
Orsay, France, in 2005. \\
\indent
He is presently research engineer at RealEyes3d, a young company 
involved in the design of camera phone applications. 
He is interested in inverse problems in image processing, and more 
specifically in improving image quality of low cost camera.
\end{biography}
\begin{biography}[{\includegraphics[width=1in,height=1.25in,clip,keepaspectratio]{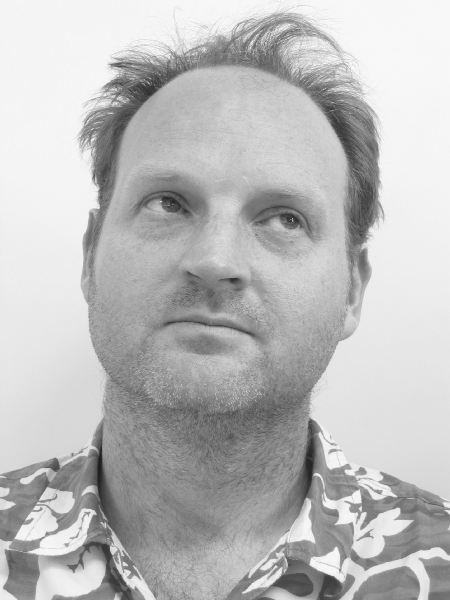}}]{Fr{\'e}d{\'e}ric Champagnat} 
was born in Dakar, Senegal, in 1966. He
graduated from the {\'E}cole Nationale Sup{\'e}rieure de Techniques
Avanc{\'e}es in 1989 and received the Ph.D. degree in physics from the
Universit{\'e} de Paris-Sud, Orsay, France, in 1993. In 1994-95 he was with
the Biomedical Engineering Institute of the {\'E}cole Polytechnique, Montreal,
Canada for a postdoctoral position. Since 1998, he is with Office National
d'\'Etudes et Recherches A\'erospatiales, Ch\^atillon, France. \\
\indent 
His main
interests are in the field of spatio-temporal processing for space or aerial
image sequences, in particular registration, motion estimation,
super-resolution and detection.\end{biography}
\begin{biography}[{\includegraphics[width=1in,height=1.25in,clip,keepaspectratio]{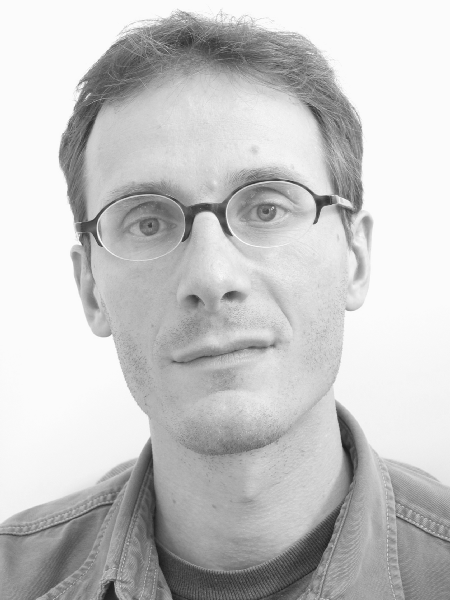}}]{Guy Le Besnerais} 
was born in Paris, France, in 1967. He graduated from
the {\'E}cole Nationale Sup{\'e}rieure de Techniques Avanc{\'e}es in 1989 and
received the Ph.D. degree in physics from the Universit{\'e} de Paris-Sud,
Orsay, France, in 1993. Since 1994, he is with Office National d'\'Etudes et
Recherches A\'erospatiales, Ch\^atillon, France. \\
\indent
His main interests are in the
fields of image reconstruction, structure-from-motion and spatio-temporal
processing for space and aerial image sequences.
\end{biography}
\begin{biography}[{\includegraphics[width=1in,height=1.25in,clip,keepaspectratio]{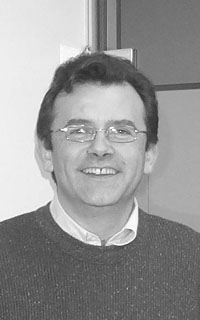}}]{Jean-Fran\c{c}ois Giovannelli} was born in B\'eziers, 
France,
in 1966. He graduated from the \'Ecole Nationale Sup\'erieure de
l'\'Electronique et de ses Applications in 1990. He received the Doctorat
degree in physics at Universit\'e Paris-Sud, Orsay, France, in 1995. \\
\indent
He is presently assistant professor in the D\'epartement de Physique at
Universit\'e Paris-Sud and researcher with the Laboratoire des Signaux et
Syst\`mes (CNRS - Sup\'elec - UPS). He is interested in regularization and
Bayesian methods for inverse problems in signal and image processing.
Application fields essentially concern astronomical, medical and geophysical
imaging. \end{biography}

\end{document}